\documentclass[
aps,prd,
showpacs,twocolumn,notitlepage,
amssymb,amsmath,amsfonts,
nofootinbib,superscriptaddress,
floats,
amsmath,amssymb,
aps,
]{revtex4-2}

\usepackage{aas_macros}
\usepackage{graphicx}
\usepackage{dcolumn}
\usepackage{bm}
\usepackage{xcolor}

\graphicspath{{./}{../pictures/}}

\newcommand{\code}[1]{\texttt{#1}}
\newcommand{\mesa}{\code{MESA} }
\newcommand{\arepo}{\code{AREPO} }

\usepackage{amsmath}	
\usepackage{amssymb}	
\usepackage{cases}
\usepackage{array,multirow}
\usepackage{upgreek}
\usepackage{textcomp}
\usepackage{lineno}
\usepackage{subcaption}

\usepackage{hyperref}
\usepackage[referable]{threeparttablex}
\usepackage{booktabs, dcolumn}

\newcommand{\Msol}{\,\mathrm{M}_\odot}
\newcommand{\Rsol}{\,\mathrm{R}_\odot}

\newcommand{\yr}{\,\mathrm{yr}}

\newcommand{\pc}{\,\mathrm{pc}}
\newcommand{\Mpc}{\,\mathrm{Mpc}}

\newcommand{\Mbh}{M_{\rm{MBH}}}
\newcommand{\sbh}{m_{\rm{sBH}}}
\newcommand{\ms}{M_{\star}}

\def\apjl{ApJL}

\def\aap{A\&A}

\def\apjs{ApJ Supp.}

\usepackage[normalem]{ulem}

\begin{document}

\title{HYDRODYNAMIC SIMULATIONS OF TIDAL DISRUPTION ENCORES}

\author{Ian P.A. Johnson}
\email{ian.p.johnson@stonybrook.edu}
\affiliation{Department of Physics and Astronomy, Stony Brook University, Stony Brook, NY 11794-3800, USA}

\author{Taeho Ryu}
\affiliation{JILA, University of Colorado and National Institute of Standards and Technology, 440 UCB, Boulder, 80308 CO, USA}
\affiliation{Department of Astrophysical and Planetary Sciences, 391 UCB, Boulder, 80309 CO, USA}
\affiliation{The Max Planck Institute for Astrophysics, Karl-Schwarzschild-Str. 1, Garching, 85748, Germany}

\author{Rosalba Perna}
\affiliation{Department of Physics and Astronomy, Stony Brook University, Stony Brook, NY 11794-3800, USA}

    \begin{abstract}
We present hydrodynamic simulations with the moving-mesh code \arepo of Tidal Disruption Encores (TDEEs)  in nuclear star clusters (NSCs).  TDEEs arise when a stellar-mass black hole (sBH) disrupts a star within the NSC, producing debris that is unbound from the sBH but remains gravitationally bound to the central massive black hole (MBH), leading to a delayed secondary flare. We find that the morphology and thermodynamics of the fallback material depend sensitively on the disruption geometry,
MBH mass, and sBH-MBH separation. We identify two distinct morphological outcomes: ring encores,  where debris
circularize into a torus,
and direct encores, where streams plunge
toward the MBH,
with encore luminosities peaking at times corresponding to the freefall timescale and one orbital period, respectively. Across all simulated cases,  we find these events exhibit luminosities of $10^{40}-10^{42}$ erg/s with lightcurves characteristic of their morphology. 
 Our work greatly improves the predictions of TDEE lightcurves and empowers observations to probe into NSC dynamics and sBH population while providing possible explanations for anomalous TDE-like flares.  
\end{abstract}


\maketitle

\section{Introduction} \label{sec:intro}

Nuclear star clusters (NSCs), made up of very dense concentrations of stars and the compact objects they leave behind, are often found in the central regions of galaxies surrounding the central massive black hole (MBH). In particular, the fraction of nucleated galaxies is found to be especially large, $\sim 90\%$, in early-type galaxies with masses $\sim 10^9 \Msol$, albeit dropping at both the high and low end of the mass distribution \citep{Sanchez2019}. The star density in NSCs is very high, possibly reaching $\sim 10^6 \Msol$~pc$^{-3}$ at a radial distance of $\sim 0.1$~pc from the core and following a very steep profile, $\propto r^{-1}-r^{-3}$ (see e.g. \citealt{Neumayer2020} for a comprehensive review). For our Galactic NSC, the stellar density is observed to be $\sim 10^6 \Msol$~pc$^{-3}$ even at a distance of 0.25 pc \citep[][]{Genzel+2010}.

With about $\sim 1\%$ 
of their mass in stellar-mass black holes (sBHs) \citep{BahcallWolf1976}, NSCs are ripe environments for close interactions between stars and sBHs. In some cases, the immense tidal forces exerted during such interactions can partially or completely disrupt the star, producing what  are known as micro tidal disruption events (TDEs) \citep{Perets2016,Lopez2019,Kremer2019,Wang2021, Ryu2022,Ryu2023a}. The prefix ``micro'' distinguishes these events from ``standard'' TDEs, which involve the disruption of stars by MBHs in galactic nuclei \citep{Hills1988,Rees1988}. 

micro-TDEs in NSCs can lead to an interesting new phenomenology, which was uncovered by \citet{Ryu2024TDEE}: For a subset of these events, the unbound debris to the sBH remains bound to the MBH, accreting at a later time, thus giving rise to a second flare, which they called a TDE Encore (TDEE). TDEE rates in galaxies with a NSC were found to vary between $\sim 10^{-6}$~yr$^{-1}$~gal$^{-1}$ for a
 MBH of $\sim 10^6\Msol$, to $\sim 10^{-3}$~yr$^{-1}$~gal$^{-1}$ for a MBH of $\sim 10^9\Msol$, if the Bahcall-Wolf density profile \citep{BahcallWolf1976} is 
 extrapolated inward toward the MBH, down to the radius where at least one star and one BH are expected to be present.

 \citet{Ryu2024TDEE} studied the evolution of the TDEE
 analytically, assuming ballistic orbits for the stellar debris. They found that the temporal evolution of the fallback rate of the debris can vary between the standard $t^{-5/3}$ power law to much steeper decays.  
 If the first flare from the micro-TDE is missed, then the TDEE can show up as an 'impostor' among the more typical TDEs from MBHs. 
 The variety of fallback rates exhibited by these encores
 may indeed provide a natural explanation for a wide range of decay slopes \citep[$t^{-1}-t^{-4}$;][]{Hammerstein+2023} of observed light curves of  optical/UV TDE candidates which are in tension with the predictions of the classical TDE model \citep[e.g.,][]{Gezari2021}. 

 Motivated by the relatively high frequency of encores per NSC, and their potential importance both as contributors to the standard MBH TDEs as well as probes of the sBH population in NSCs \citep{Fragione2021}, here we present hydrodynamic simulations of TDEEs, performed using the moving-mesh code 
\arepo\ \citep{Arepo,ArepoHydro,Arepo2}, to investigate debris morphology and observables, accounting for non-linear hydrodynamical effects.

We find that TDEEs are broadly described by two morphologies which we call ``direct encores" and ``ring encores",  which display distinct observational signatures. In direct encores, the debris following the initial micro-TDE 
undergoes collisions with other streams, creating shocks on the freefall timescale and the resulting radiation is emitted on the photon diffusion time.
 For ring encores, the debris begins to reheat after a single period $\mathcal{P}$ of the sBH's orbit around the MBH and forms a mostly uniform ring after $\approx 4\,\mathcal{P}$. Rapid cooling then follows. Based on viscous timescale estimates, debris will make it into the relativistic regime of the MBH on the timescale of weeks to months for direct encores and years to decades for debris rings which would then produce a rebrightening in luminosity.

In Section \ref{sec: analytic expec}, we review the analytic theory  of TDEEs outlined in \cite{Ryu2024TDEE} and describe the different morphology types. Section \ref{sec: num methods} details our use of the \arepo and \mesa codes, along with the post-process analysis. We present the results for direct encore flares in Section \ref{sec: direct encore results}, followed by the simulations of ring encores in Section \ref{sec: ring encore results}. Finally, 
in Section \ref{sec: discussion},  
we discuss the observational implications and the broader astrophysical context. 
We summarize and conclude in Section \ref{sec: Conclusion}.

\section{Analytic Expectations - Direct Encore or Ring?}
\label{sec: analytic expec}
\citet{Ryu2024TDEE} discussed how the unbound debris from micro-TDEs, which constitute approximately half the original stellar mass, can orbit within a larger gravitational well of a nearby MBH. To qualitatively discuss the final fate of the unbound debris ``spilled'' over to the MBH, it is convenient to introduce the characteristic orbital-energy width of debris produced in a micro-TDE relative to the sBH,
\begin{align}
    \Delta\epsilon=\frac{G\sbh R_{\star}}{R_{\rm t}^{2}},
\end{align}
where $R_{\rm t}=(\Mbh/M_{*})^{1/3}R_*$ is the tidal radius and $M_{*}$ ($R_{*}$) is the stellar mass (radius). If the original orbit of the star is parabolic relative to the sBH, the spilled debris' orbital energy relative to the MBH can be approximated as $E\simeq\epsilon_{\rm sBH}+\Delta \epsilon$, where $\epsilon_{\rm sBH}=-G\Mbh/2R_{\rm bin}$ is the orbital energy of the sBH on a circular orbit relative to the MBH when their distance is $R_{\rm bin}$. 

We can thus consider three possible scenarios:\\ 1) If $|\epsilon_{\rm sBH}|\gg\Delta \epsilon$ (or $E\simeq\epsilon_{\rm sBH}$), the energy change due to the micro-TDE is negligible and the spilled debris will essentially share the same circular orbit as the sBH. However, because of the internal energy spread within the debris, there will be a redistribution along the circular path, ultimately leading to a ring. \\
2) If $|\epsilon_{\rm sBH}|$ is comparable to, yet larger than $\Delta \epsilon$, the spilled debris remain bound ($E<0$). In this case, $E$ can be significantly smaller than either of $\Delta \epsilon$ or $\epsilon_{\rm sBH}$, and the debris may follow a highly eccentric orbit around the MBH, falling into it on a free-fall timescale. \\
3) If $|\epsilon_{\rm sBH}|<\Delta \epsilon$ ($E>0$), the spilled debris are also unbound from the MBH and escape the binary black hole system. 

\begin{figure}
    \centering\includegraphics[width=\linewidth]{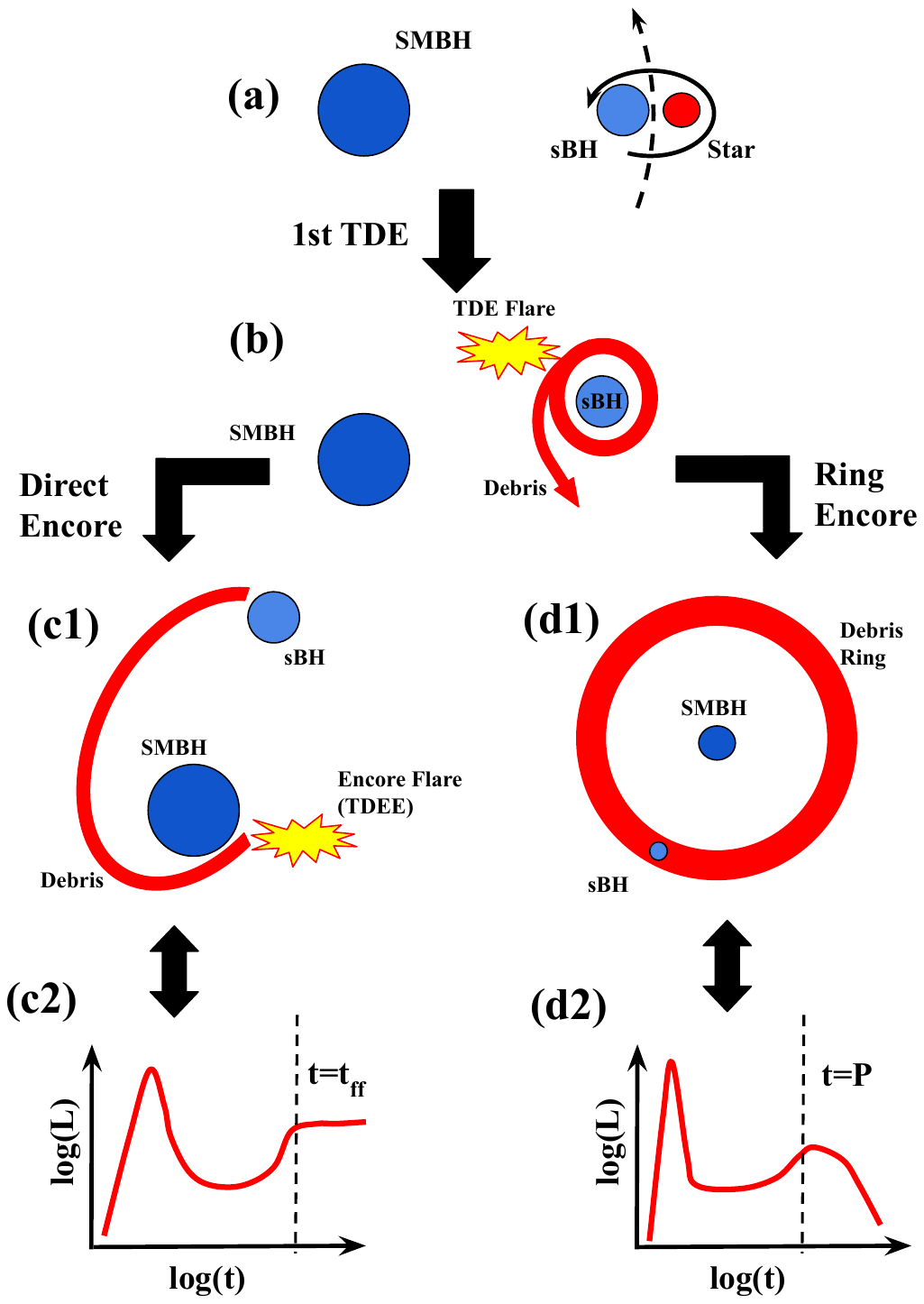}
    \caption{Overview of the morphology types of TDEEs. (a) Initial three-body system of an MBH orbited by a sBH-star binary. (b) Star undergoes a micro-TDE onto the sBH while orbiting the MBH. Post micro-TDE, a direct encore (c1) or a debris ring (d1) ensue, with corresponding schematic lightcurves displayed in (c2) for a direct encore and (d2) for a ring. }
    \label{fig:overview_fig}
\end{figure}

To summarize, following the first TDE flare (Fig. \ref{fig:overview_fig}a), the debris that do not fall into the sBH have three possible fates:
\begin{enumerate}
    \item Falling into the MBH on a dynamical timescale and creating a second flare (Fig. \ref{fig:overview_fig}c).
    \item Retaining their original trajectory and forming a ring which will accrete onto the MBH on a viscous timescale (Fig. \ref{fig:overview_fig}d).
    \item Escaping the binary BH system (not pictured).
\end{enumerate}


Since direct encores follow ballistic debris trajectories, the closer the initial TDE is to the MBH, the more likely for a secondary flare. The angle $\varphi$ subtended by the innermost stable circular orbit (ISCO) for massive particles around a Schwarzchild MBH is given by
\begin{equation}
    \varphi = 
    2\arcsin\left(\frac{6r_\text{g}}{ R_\text{bin}}\right)\approx
    \frac{12 GM_\text{MBH}}{ R_\text{bin}c^2}\,,
\end{equation}
where the approximation is for small angles $\varphi$ and $r_\text{g}=GM_\text{MBH}/c^2$ is the gravitational radius of the MBH.

\begin{figure}
    \centering
    \includegraphics[width=\linewidth]{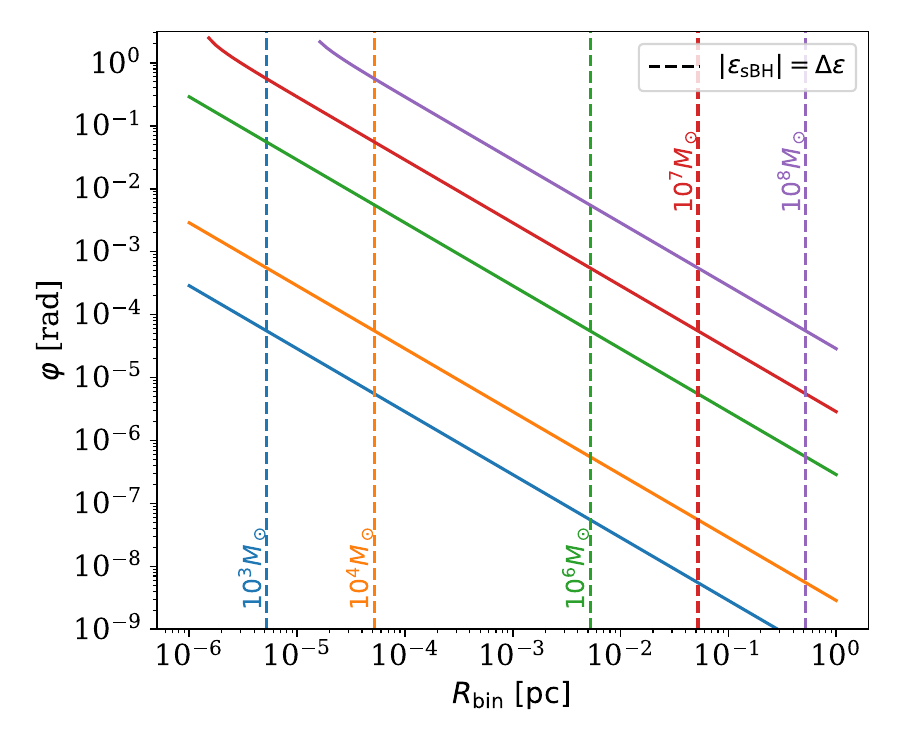}
    \caption{{ The angle $\varphi$ subtended by the MBH's ISCO versus the radial separation $R_\mathrm{bin}$ for a variety of MBH masses.}
    Rings are likely to form to the left of the dashed vertical lines, where $|\epsilon_\mathrm{sBH}|>\Delta\epsilon$. To the right of the lines, direct path encores are possible depending on the specific orbital configuration of the three bodies.}
    \label{fig:subtended_angle}
\end{figure}

The angle subtended is not intended to be an exact expression for the likelihood of the debris being disrupted again by the MBH since the width of the debris, gravitational attraction of the MBH, and existence of less direct debris trajectories all improve the chances of an encore. In part, this motivates our use of hydrodynamic simulations of the phenomenon across several initial angles. Nonetheless, the scaling displayed in Figure \ref{fig:subtended_angle} is a helpful guide to the family of TDEE flares.

\section{Numerical Methods}
\label{sec: num methods}

\subsection{Stellar model}
We created a stellar model assuming near-solar metallicity ($Z=0.02$) using the stellar evolution code \mesa\ \citep[version r24.03.1;][]{Paxton+2011,paxton:13,Jermyn+2023}. We followed the prescriptions for overshoot, convection, semiconvection, and thermohaline mixing from \cite{Choi+2016}, and employed the Ledoux criterion \cite{ledoux_stellar_1947} to identify the convective region boundary. We evolved a main-sequence star with a mass of $1\Msol$ up to when the central hydrogen mass fraction reached 0.3. At that time, the star's surface radius is $1\Rsol$ as anticipated. We show the density and Hydrogen mass fraction profiles of the stellar model in 
Appendix~B.

\begin{figure*}
    \centering
    \begin{subfigure}[t]{0.62\textwidth}
        \centering
        \includegraphics[width=\linewidth]{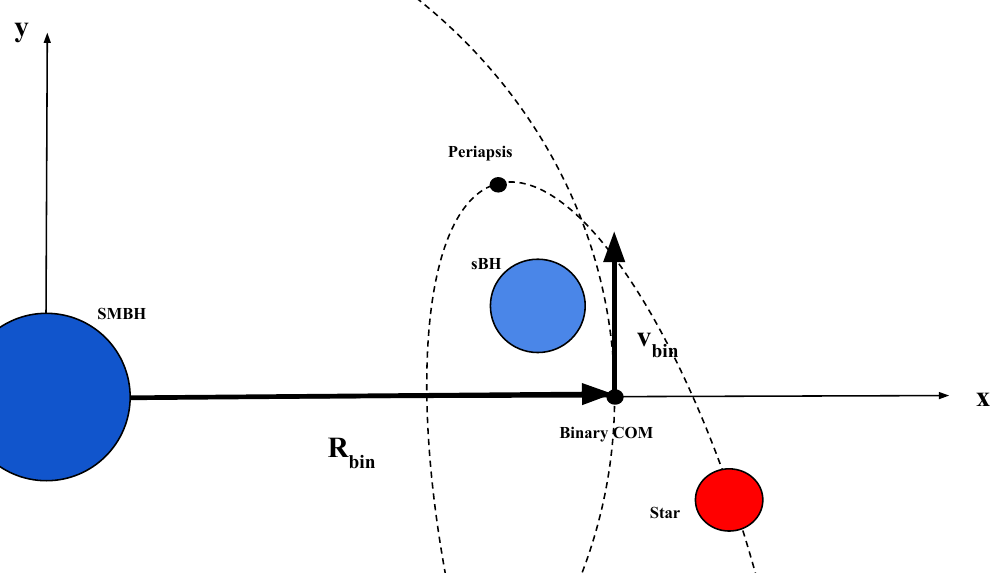}
        \caption{}
        \label{fig:TDEE_diagram_a}
    \end{subfigure}
    \hfill
    \begin{subfigure}[t]{0.36\textwidth}
        \centering
        \includegraphics[width=\linewidth]{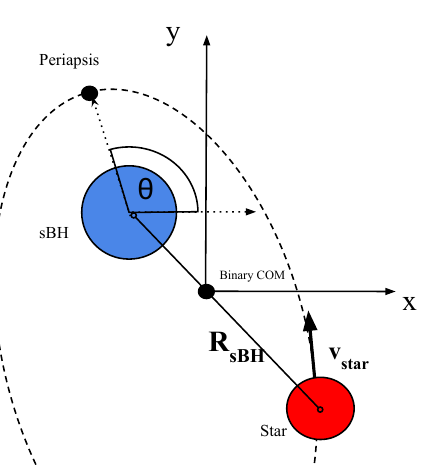}
        \caption{}
        \label{fig:TDEE_diagram_b}
    \end{subfigure}
    \caption{Schematic depiction of the initial 3 body setup, not to scale: (a) Full 3-body system in the MBH rest frame; (b) Zoom on the sBH-star binary in the sBH rest frame.
    }
    \label{fig:TDEE_diagram}
\end{figure*}

In this first hydrodynamic investigation of TDEEs, we restrict our simulations to Sun-like stars. Although quantitative details such as energetics and timescales will depend on stellar mass, the main qualitative trends of TDEEs identified here are expected to remain robust. Fixing the stellar mass allows us to isolate and study the dependence on encounter geometry, MBH mass, and sBH–MBH separation.

\subsection{Hydrodynamic simulations}

We performed a suite of 3D hydrodynamic simulations of TDEEs using the moving-mesh magnetohydrodynamics code \arepo\ \citep{Arepo,ArepoHydro,Arepo2}. We employed an ideal gas equation of state accounting for radiation pressure assuming local thermodynamic equilibrium. 

We first mapped a 1D \mesa\ profile onto a 3D \arepo\ domain using the method by \citet{Ohlmann+2017}, and then relaxed the star using $N=5\times 10^5$ cells. This resolution was chosen after performing convergence tests for key quantities such luminosity, as discussed in 
Appendix~A of Section~\ref{sec: res tests}.

We initialized the simulation with the $1 \Msol$ star and two non-spinning point mass particles with masses $\Mbh$ and $\sbh$ representing the MBH and sBH, respectively. These BH particles have no intrinsic angular momentum,
interact with gas only via gravity, and have separate softening lengths based on the system's scale. 

In this work, we adopt the 2+1 body limit of the three-body problem to initialize the system. In this approximation, the dynamics are decomposed into a sub-binary, comprised of the sBH and the star, and an outer orbit in which the MBH moves relative to the center of mass (COM) of the sub-binary. This hierarchical treatment assumes a large separation between the MBH and the sBH–star pair compared to the binary’s internal separation, such that the gravitational influence of the MBH acts primarily as an external perturbation on the sub-binary. The 2+1 body limit thus allows the complex three-body configuration to be approximated as a coupled system of two Keplerian orbits: one describing the motion of the star about the sBH, and another describing the orbital motion of the sub-binary COM around the MBH.

All gas cells are initialized with identical initial velocities derived from this configuration. The sBH–star system is treated as a Keplerian sub-binary, while the MBH is set in a coplanar, prograde circular orbit around the COM of the sub-binary. The velocity field of the simulation is constructed by superposing the velocity contributions from the internal sub-binary motion and the outer orbital motion of the sub-binary COM about the MBH. All simulations are carried out in the center-of-mass frame of the full three-body system, which approximately coincides with the MBH position due to its dominant mass. A schematic representation of this 2+1 body setup is shown in Fig.~\ref{fig:TDEE_diagram}, illustrating the hierarchical orbital structure and relative geometry of the system.

The specifics of the sBH-star binary orbit are held constant across the various simulations, except for the rotational orientation $\theta$, defined as the predicted angle at which the periapsis will occur. Owing to three body dynamics, this prediction is not exact, but it provides a good approximation in the 2 + 1 body limit. 

Parameters of the sBH-star binary which are held constant are the masses, the eccentricity $e$, the initial separation $R_\mathrm{sBH}=2\,R_\mathrm{t}$ (where  $R_\mathrm{t}$ is defined with $\sbh$ and $\ms$), and the distance to periapsis $r_\mathrm{p} = 0.25 \, R_\mathrm{t}$.  We choose as fiducial values $M_{\star}=1\Msol$, $\sbh=10\Msol$ and $1-e=10^{-3}$. 

In total, we performed 8 simulations in addition to those required for resolution testing. We detail in Table~\ref{tab:merged_simulations} the variable input simulation parameters, together with a key output for each, that is the (approximate) peak encore luminosity, computed as described in the following.

\begin{table}[h]
    \centering
    \begin{tabular}{cccc}
        \specialrule{1pt}{0pt}{0pt} 
        \multicolumn{4}{c}{\textbf{TDEE Simulations}} \\
        \specialrule{1pt}{0pt}{0pt} 
        $M_{\text{MBH}}$ ($M_{\odot}$) & $R_{\text{bin}}$ [pc] & $\theta$ [Rad] & $L_\mathrm{enc}$ [erg/s]\\ \midrule
        \multicolumn{4}{c}{\textit{Direct Encore Outcomes}} \\ \midrule
         $10^3$ & $10^{-5}$ & $0$ & $10^{40}$\\
         $10^3$ & $10^{-5}$ & $\pi/4$ & $10^{40}$\\
         $10^3$ & $10^{-5}$ & $\pi/2$ &  $10^{41}$\\
         $10^3$ & $10^{-5}$ & $3\pi/4$ & $10^{41}$ \\
         $10^3$ & $10^{-5}$ & $\pi$ & $10^{41}$\\
         $10^3$ & $10^{-5}$ & $3\pi/2$ & $10^{40}$\\ \midrule
        \multicolumn{4}{c}{\textit{Ring Encore Outcomes}}\\ \midrule
         $10^6$ & $10^{-5}$ & $3\pi/4$ & ${}^*10^{42} $ \\
         $10^6$ & $10^{-4}$ & $3\pi/4$ & $10^{40}$\\ \midrule
    \end{tabular}
    \caption{ List of  the TDEE simulations. From left to right: MBH mass $\Mbh$, separation $R_{\rm bin}$ between sBH and MBH, argument of periapsis $\theta$,  approximate peak luminosity of the second flare $L_{\rm enc}$. 
    $^*$For sufficiently compact ring formation, the distinction between micro-TDE and encore flares may become blurred.
    }
    \label{tab:merged_simulations}
\end{table}

\begin{figure*}
    \centering
    \includegraphics[width=.95\linewidth]{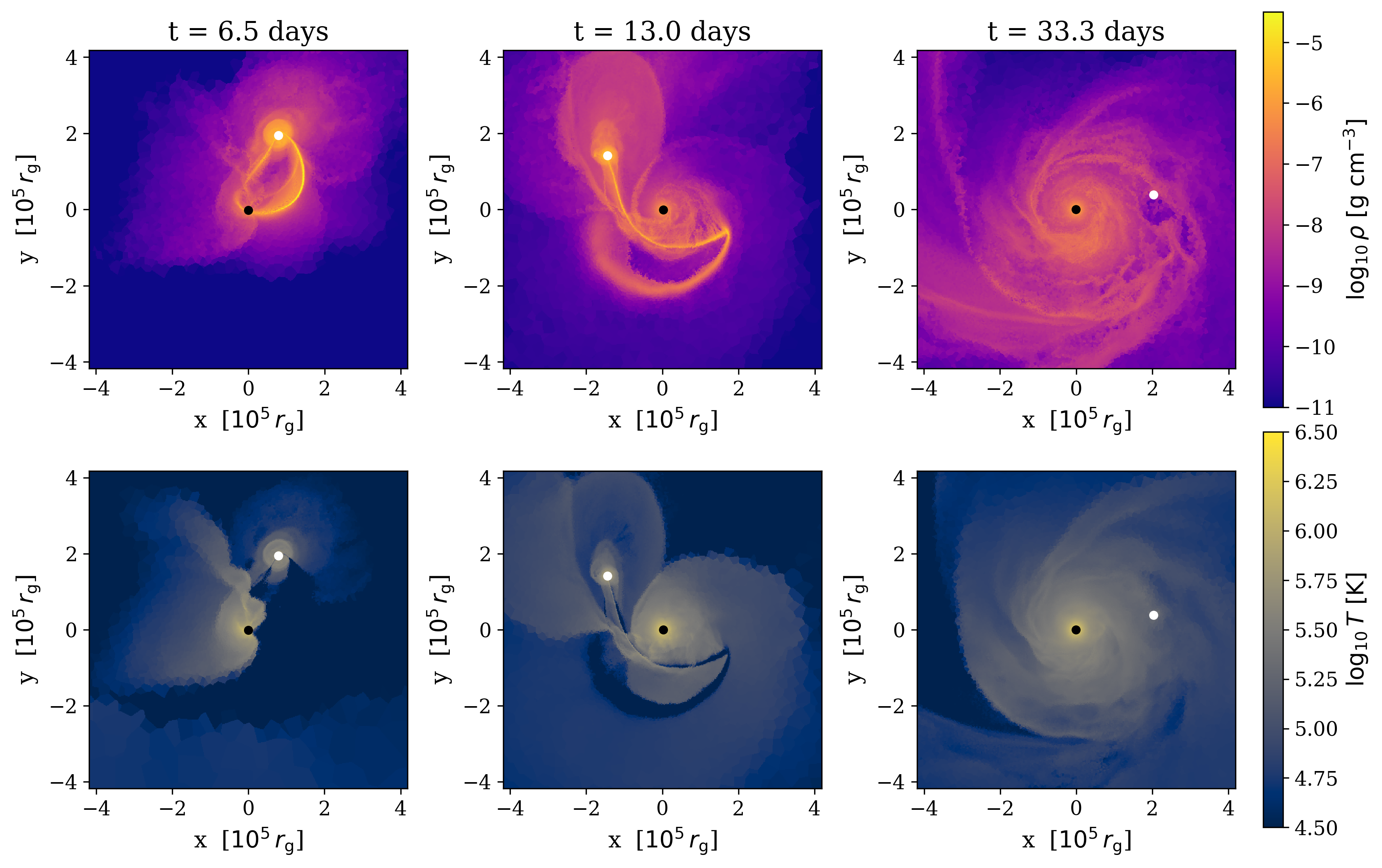}
    \caption{Density (top) and temperature (bottom) evolution in a simulation of a direct encore  with $\Mbh=10^{3}\Msol$ (black dot in the center), $\sbh=10\Msol$ (white dot), $R_{\rm bin}=10^{-5}\pc \simeq 2\times 10^{5}r_{\rm g}$, and $\theta=3\pi/4$. This is the highest luminosity encore we simulated, but it shares many features with other members with the same TDEE morphology. After a time $t\sim t_\mathrm{ff}\approx6\;\mathrm{days}$, the star is destroyed by the sBH, resulting in an accretion flow near the sBH and a stream falling into the MBH. After a time $t\sim 2t_{\rm ff}$, the tidal tail folds on itself, followed by more circular flows spreading even out to the distance of the sBH.}
    \label{fig:direct encore evo}
\end{figure*}

\subsection{Bolometric Luminosity} 
\label{Bolometric Luminosity from AREPO}

Since our simulations did not include explicit radiation-transfer calculations, we estimated the luminosity from the radiation-pressure-dominated, optically thick debris in post-processing,  following \citet{Ryu+2023b,Ryu+2023,Ryu+2024b}. We first constructed a cylindrical ($r,\phi,z$) virtual grid centred at the sBH with an exponential radial grid extending from $R_\epsilon$ to $3R_\mathrm{bin}$ (i.e. $\Delta r / r\simeq \log_{10}(3R_\mathrm{bin}/R_\epsilon)/N_r\simeq 0.016$ where $R_\epsilon=10^{-3}\mathrm{cm}$), a uniform azimuthal grid ($\Delta \phi=2\pi/N_\phi=0.044$), and an exponential vertical grid ($\Delta z/z \simeq 0.016$). The total bolometric luminosity is obtained by integrating the radiative flux $F_{\rm rad}(t,r,\phi)$ through each vertical column: 
\begin{align}
    L(t) = \int_{R_\epsilon}^{3R_\mathrm{bin}}dr\int_{0}^{2\pi} d\phi F_{\rm rad}\,.
\end{align}
Considering the symmetry around $z=0$, we explicitly calculated only the upward flux relative to the midplane. Downward flux was also calculated and found to be nearly identical verifying our symmetry assumptions. We computed the upward flux as:
\begin{align}
    F_{\rm rad}(r,\phi)&=\frac{\int_{z_{\tau}}^{z_{\rm ph}} dz aT^{4}}{t_{\rm diff}}\,,
    \label{eq: Frad}
\end{align}
where ${z_{\rm ph}}$ is the height above which $t_{\rm diff}<t$, with $t$ being the age of the system, and $z_{\tau}$ is the height of the photosphere at $\tau\simeq 1$. 
Because only photons with $t_{\mathrm diff}<t$  can reach and escape the photosphere, we restricted the vertical integration to the range between ${z_{\mathrm ph}}$ and ${z_{\mathrm max}} = 3R_\mathrm{bin}$. To ensure that only radiation from the gas originating from the star is included, we applied a density cut above $10^{-15} \mathrm{g/cm^3}$ \footnote{Density is chosen such that the cut-off density is above the so called vacuum cells initialized but 2-3 orders of magnitude below the density of the main ring structure.} when calculating the thermal energy within each column. We calculated the optical depth as
\begin{align}
    \tau(z) = -\int_{z_{\rm max}}^{z} \kappa \rho dz,
\end{align}
where the opacity $\kappa$ was obtained from the OPAL Rosseland mean opacity table \citep{IglesiasRogers1996} for given local density $\rho$ and temperature at each height. We ensured that $z_{\rm max}$ is sufficiently high to enclose the photosphere within the virtual grid. Accordingly, the photon diffusion time is
\begin{align}
    t_{\rm diff}(t,r,\phi)\simeq\frac{H\tau(z=0)}{c},
    \label{eq:tdiff}
\end{align}
where $H$ is the vertical density-weighted scale height at the given $r$ and $\phi$. The diffusion time represents an ``average'' timescale across the entire vertical column, in the sense that it does not depend on height. While this is admittedly an approximate estimate, it is robust against uncertainties in the exact photosphere location, which is a limitation inherent to estimates based on adiabatic hydrodynamical simulations. { In Appendix~\ref{sec:SEDs}, we then expand on the results presented here and use our best estimate of the photospheric location in order to compute the spectral energy distributions associated with different TDEE morphologies. }

\subsection{Viscous Timescale}
\label{sec: t_visc}

The viscous timescale is calculated using the $H(r,\phi)$ values from our cylindrical virtual grid, adopting the $\alpha-$prescription \citep{Shakura1973}:
\begin{equation}
 t_\mathrm{visc}(r,\phi)\approx\frac{1}{\alpha}\left(\frac{r^3}{G \Mbh}\right)^{1/2} \left(\frac{H(r,\phi)}{r}\right)^{-2}\,,   
 \label{eq:tvisc}
\end{equation}
where we take $\alpha=0.1$. The mass distribution of the viscous timescale is calculated over $r$ and $\phi$ coordinates for masses $M(r,\phi)$ and the corresponding $t_\mathrm{visc}(r,\phi)$. For the purposes of Fig.~\ref{fig:direct_timescales_visc}, we average $t_\mathrm{visc}(r)$ over $\phi$, weighted by mass, that is $t_\mathrm{visc}(r) = \sum_\phi (t_\mathrm{visc}(r,\phi) \cdot M(r,\phi))/\sum_\phi M(r,\phi)$, where $M(r,\phi)$ is the cell mass at given $r$ and $\phi$. 

\section{Direct Encore Results}
\label{sec: direct encore results}

In one limit of TDEEs,
the debris has an orbital energy relative to the MBH that is negative but close to zero.
Consequently, the debris trajectory can be strongly perturbed by the initial micro-TDE and redirected onto a highly eccentric orbit that plunges toward the MBH.

We performed simulations with $\Mbh=10^3\Msol$ and $R_\mathrm{bin}=10^{-5} \pc$ 
exploring a range of sBH–star orientations characterized by the projected periapsis angle $\theta$. We focused on varying $\theta$ to assess how sensitively encore formation depends on the debris orientation.

We note that for our direct encore simulations, the MBH is best considered an intermediate-mass black hole or IMBH. Owing to computational limitations, a direct encore simulation for more typical SMBH masses is not feasible; however, MBHs of $10^3\Msol$ are still valuable to explore the overall dynamics of TDEEs and are of worthwhile discussion on their own (more in Section \ref{sec: discussion IMBH}). 

We show three snapshots of the TDEE
simulation with $\theta=3\pi/4$ 
in Fig.~\ref{fig:direct encore evo}.
The top panels show the debris density, while the bottom ones the temperature.
Past analytical models of \citet{Ryu2024TDEE} envisioned debris of the micro-TDE as having direct ballistic paths to the MBH; however, Figure \ref{fig:direct encore evo} displays nuanced hydrodynamics at play. Of great relevance to the TDEE phenomenology is the wake trailing the path of the sBH. This wake contains an appreciable fraction of the total debris mass (typically $\approx70\%$ of $0.5\ms$); it trails the sBH and is prone to freefall toward the MBH, creating shocks when supersonically-moving streams intersect each other. 

\begin{figure}
    \centering
    \includegraphics[width=0.5\textwidth]{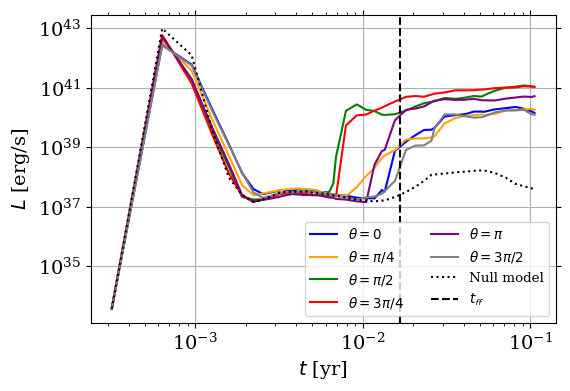}
    \caption{Bolometric luminosity of the direct encore TDEEs for different initial debris orientations. The first peak corresponds to the micro-TDE around the sBH. Shocks corresponding to the encore begin at roughly the freefall timescale, which is marked. 
    The first data point is manually added as $1 L_\odot$ at early times,
    since our virtual grid struggles to resolve an undisrupted star.}
    \label{fig:direct_luminosities}
\end{figure}

We find that, while the orientation parameter $\theta$ of the debris significantly affects their dynamics,  the peak encore luminosity varies by less than an order of magnitude as shown in Figure \ref{fig:direct_luminosities}.
The primary cause for this variation is due to the amount of mass which falls onto the MBH following the micro-TDE. 
 Figure \ref{fig:direct encore evo} displays the debris tail which contains approximately half of the total stellar mass on a direct path, but other orientations $\theta$ have longer trajectories and thus increase in luminosity later. The simulation with $\theta=\pi/2$ is a special case where the stream self-interacts early and thus the luminosity has a local peak at $t\sim 8\times 10^{-3}\yr$. Other angles outside of $\pi/2<\theta<\pi$ do not
display significant tidal tails;
 therefore, their luminosity is an order of magnitude lower. Shock-heating of the wake during freefall to the MBH further produces an observable luminosity $\approx10^{40}$ erg/s. 

We report the highest encore luminosity of $\approx 10^{41}$ erg/s, corresponding to roughly
the Eddington limit for a $10^3 M_\odot$ MBH, and the qualitatively most complex debris trajectories for $\pi/2 \leq \theta\leq \pi$. These TDEEs correspond to  disruptions occurring closer to the MBH. Although our angle $\theta$ has a slightly different definition than that of \citet{Ryu2024TDEE}, we confirm the most dramatic plunging angle for the debris to be $\theta={3\pi}/{4}$. 
\textbf{We remark that,}
because our version of \arepo does not include radiation transfer, the late time luminosity \textbf{change} of our TDEE cannot be accurately predicted with our simulations. { We restrict radiative losses - estimated via the time integrated luminosity - to  $\lesssim 5\%$ of the sytem's total thermal energy, to ensure that the hydrodynamical evolution remains robust regardless of whether radiative cooling is explicitly included.}

\begin{figure}
    \centering    \includegraphics[width=\linewidth]{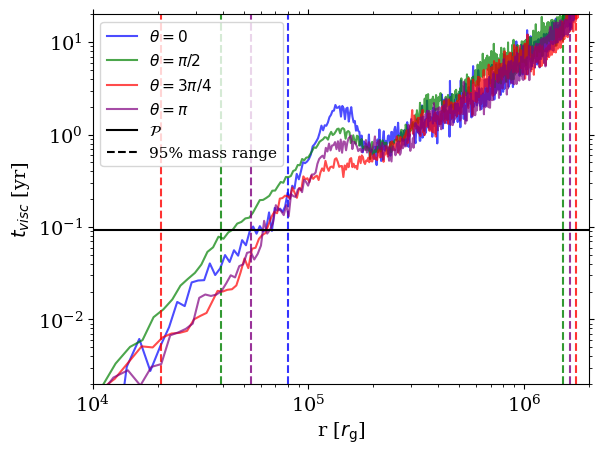}
    \caption{
    { Distribution of viscous timescales (Eq. \ref{eq:tvisc}) with respect to the debris location $r$ for a subset of direct encore simulations.} The dashed lines, sharing the same colors, indicate the range within which 95\% of the debris mass is enclosed. { We may anticipate a moderate enhancement of the accretion rate onto the MBH within the approximate timescale of the optical/UV encore flare.}}
    \label{fig:direct_timescales_visc}
\end{figure}

We note that our simulation reaches a rotationally stable disk by the end of an sBH orbital period. This final state motivates an understanding of late timescale TDEEs as being fuelled via viscous transfer of angular momentum.
In Fig.~\ref{fig:direct_timescales_visc}, we show the viscous timescale distribution for the encore debris as a proxy for the time they would reach the MBH.  Accretion of this debris to the MBH can yield an additional source of luminosity on the freefall timescale. 
As an order of magnitude estimate, we can write
$L_{\rm acc} \sim  5\times 10^{44} {\rm erg~s^{-1}}(\eta/0.1)(M_{\rm debris}/0.1M_\odot)(t_{\rm visc}/1{\rm yr})^{-1}$, where $\eta$ is an accretion efficiency factor. Accretion of the debris onto the MBH, depending on the efficiency, can therefore result in an
hyper-Eddington accretion flow
for $M_{\rm MBH}=10^3\Msol$,
possibly driving an outflow/jet, and yield a rather brilliant encore flare.

\begin{figure*}
    \centering
    \includegraphics[width=\textwidth]{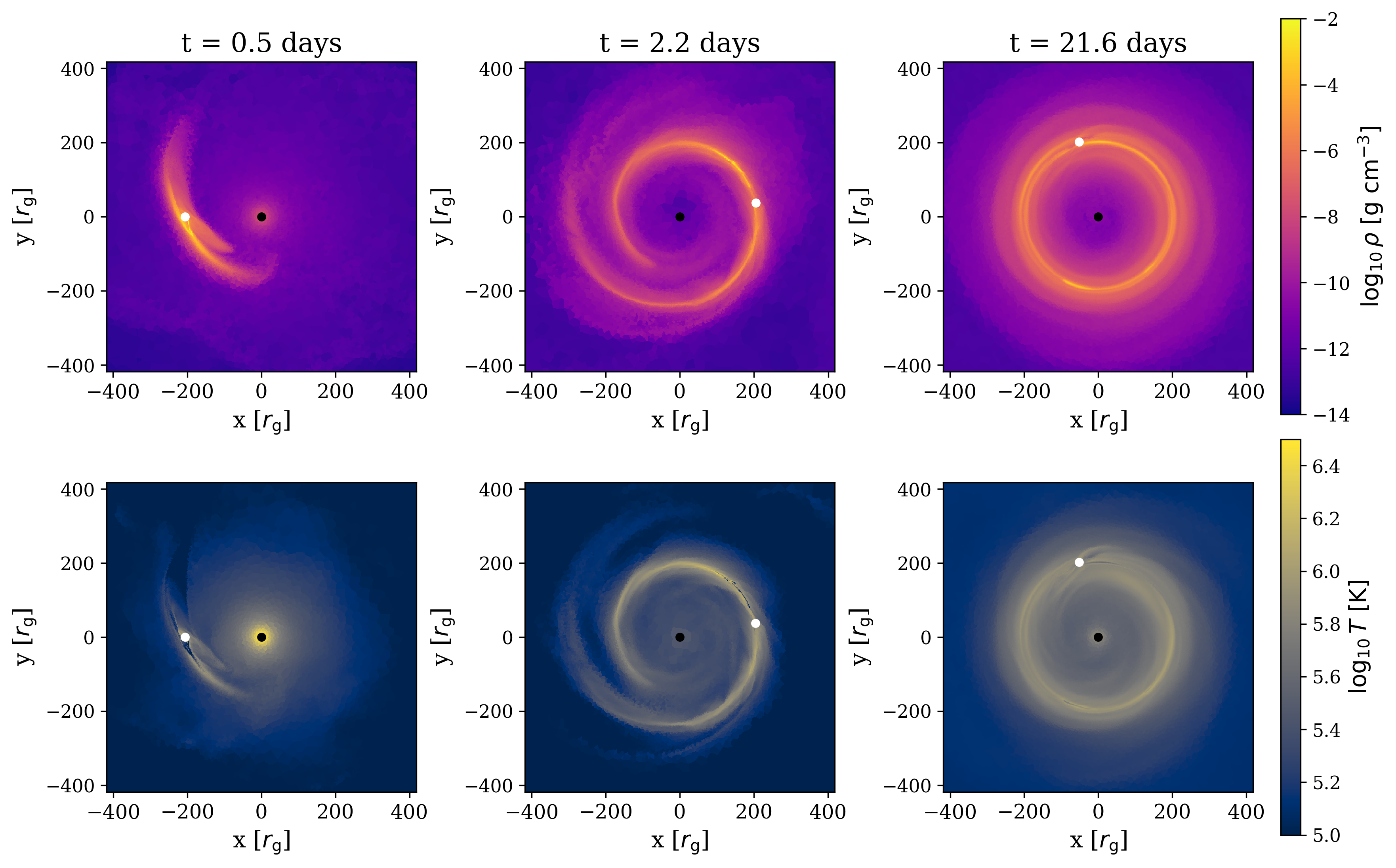}
    \caption{Disk development and evolution around a central $10^6 \Msol$ MBH (black dot at x=y=0) following a TDE of a $1\Msol$ star by a $10\Msol$ sBH (white dot). For this simulation, $\mathcal{P}\approx 1.1 \mathrm{\;days}$. The sBH-star sub system is initially set in a circular orbit around the MBH with initial separation $R_\mathrm{bin}=10^{-5}\pc=209 \,r_\mathrm{g}$. {\em Top panels:} Evolution of the disk's density; {\em Bottom panels:}  Evolution of the disk's temperature.}
    \label{fig:disk_evolution}
\end{figure*}

\section{Ring Encore Results}
\label{sec: ring encore results}

This scenario leaves the orbit of the stellar debris largely unchanged following the initial TDE flare, since their energy spread due to the micro-TDE is not large enough to significantly alter their path with respect to the initial trajectory of the star. It then follows that the debris form a disk in the shape of the original orbit as hypothesized by \citet{Ryu2024TDEE}.

Although the formation of a ring can be anticipated from an analytic comparison between $\epsilon_{\rm sBH}$ and $\Delta \epsilon$, such analytic estimates are insufficient to examine the detailed properties of the ring, due to the complex hydrodynamics and 3-body dynamics involved. Using our simulations, however, we can measure the thermodynamic properties of the ring debris, including their scale height and viscous timescale, as well as their observables, such as the bolometric luminosity.

We confirmed the formation of a ring from the debris in two simulations with $R_\mathrm{bin}=10^{-5}\text{ pc}$ and $R_\mathrm{bin}=10^{-4}\text{ pc}$. All other parameters were kept the same: $\sbh=10\Msol$, $\Mbh=10^{6}\Msol$, and $M_*=1\Msol$. For these parameters, $\Delta \epsilon/\epsilon_{\rm sBH}\simeq 10^{-2} (R_{\rm bin}/10^{-4}\pc)\ll 1$, which, according to our analytic estimates (see \S~\ref{sec: analytic expec}), corresponds to this ring-encore scenario.

As shown in Figure \ref{fig:disk_evolution} for $R_{\rm bin}=10^{-5}\pc$, the debris created in the micro-TDE closely track the orbit of the sBH. However, because of the energy spread within the debris ($\simeq \Delta \epsilon$), each gas parcel follows a slightly different orbit, producing a crescent-shaped structure (left panels) that gradually evolves into an inspiral (middle panels), and eventually merges into an axisymmetric ring (right panels) over the course of roughly 20 orbital periods of the sBH. The structural transformation of the debris from a crescent to a ring qualitatively resembles the debris evolution in ``extremely relativistic'' TDEs \citep{Ryu+2023b} (see their Fig. 1), although the two events are entirely distinct in terms of their formation mechanisms, spatial and temporal scales, subsequent evolution of the ring, and resulting observables.    

In the two subsections below, we focus on discussing relevant time and spatial scales for the ring (\S~\ref{subsec:ring_time}) and its luminosity (\S~\ref{subsec:ring_L}).

\begin{figure*}
    \centering
    \includegraphics[width=\textwidth]{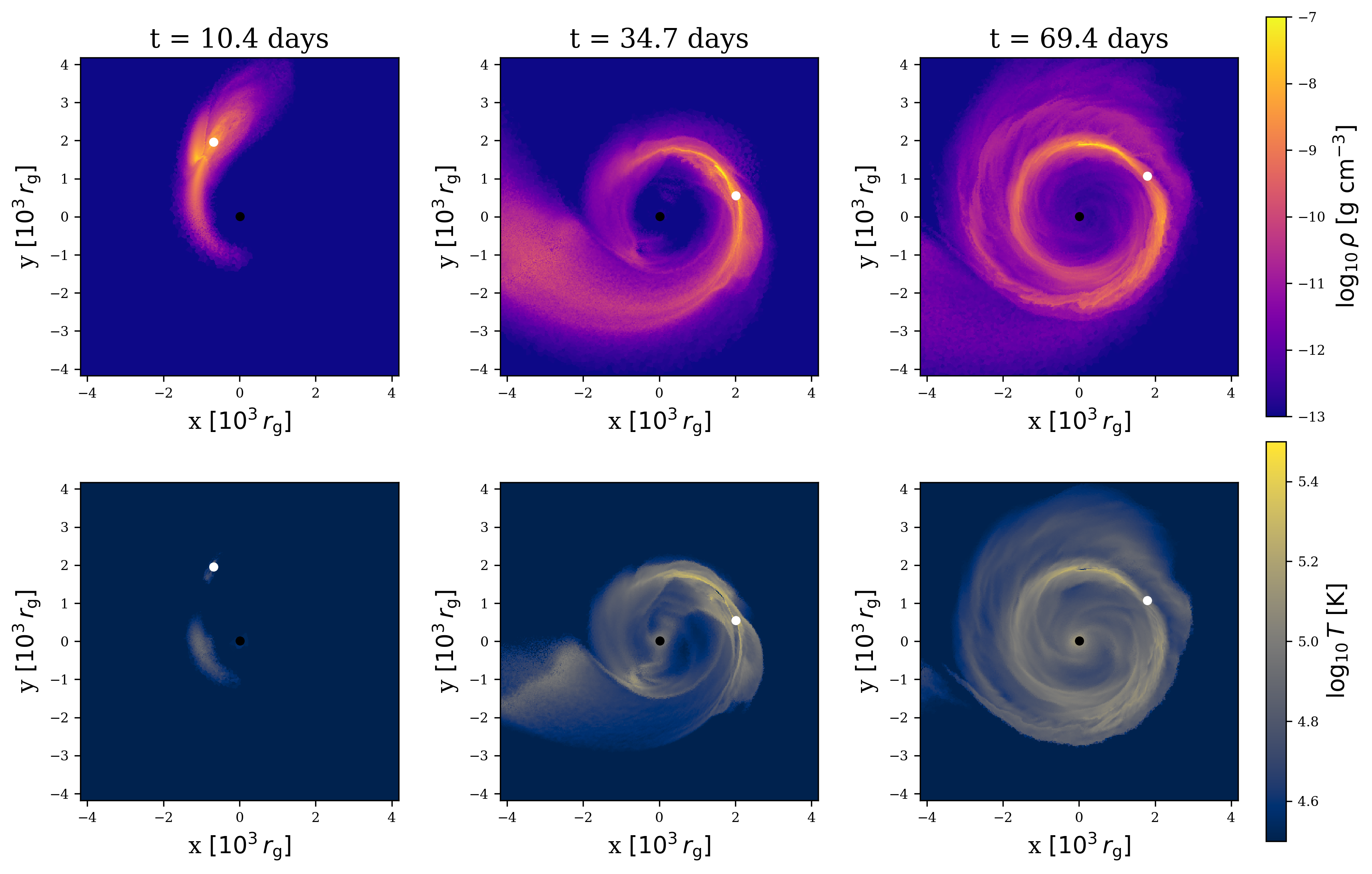}
    \caption{Disk development and evolution around a central $10^6 \Msol$ MBH (black dot at x=y=0) following a TDE of a $1\Msol$ star by a $10\Msol$ sBH (white dot). For this simulation, $\mathcal{P}\approx 34 \mathrm{\;days}$. The sBH-star sub system is initially set in a circular orbit around the MBH with initial separation $R_\mathrm{bin}=10^{-4}\pc=2.1\times10^3 \,r_\mathrm{g}$. {\em Top panels:} Evolution of the disk's density; {\em Bottom panels:}  Evolution of the disk's temperature.}
    \label{fig:disk_evolution_10x}
\end{figure*}

\begin{figure}
    \centering
    \includegraphics[width=0.45\textwidth]{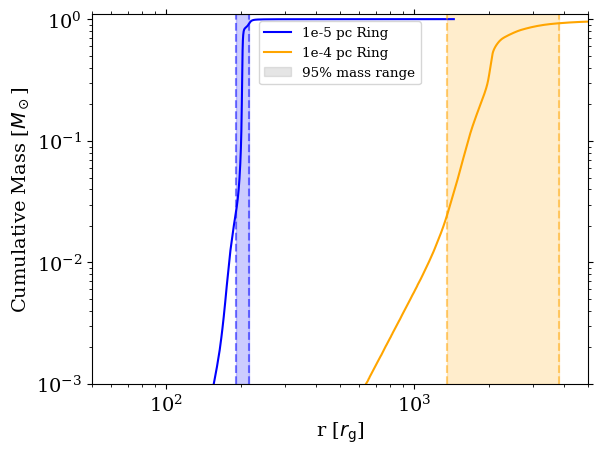}
    \caption{{ Cumulative radial mass distribution of TDEE rings, evaluated after the torus formation at $t\approx5\mathcal{P}$. 
    The shaded regions indicate
    the radial extent of the debris enclosing
$95\%$ of the debris mass, corresponding to $\sim 26r_\mathrm{g}$  
    for the $10^{-5}$ pc ring and $\sim 2.5 \times 10^3r_g$ for the ring centered at $10^{-4}$ pc.}}
    \label{fig:ring_width}
\end{figure}

\subsection{Width and Timescale of the Ring}\label{subsec:ring_time}

Our results indicate a disk aspect ratio of $H/R_\mathrm{bin}\approx10^{-2}(R_{\rm bin}/10^{-4}\pc)$, so $H\propto R_{\rm bin}^{2}$. On the other hand, the radial extent of the ring $\Delta r$, defined as the width enclosing 95\% of the total mass in the ring, is approximately $\Delta r/R_{\rm bin}\simeq 65(\Delta\epsilon / \epsilon)$ so $\Delta r\propto R_{\rm bin}^{2}$. This yields $H/\Delta r\simeq 1.5\times10^{-3}$, indicating that the vertical cross-sectional shape of the axisymmetric debris is more extended in the radial direction. The radial extent of the debris in the two simulations for the ring encore scenario is illustrated in Fig.~\ref{fig:ring_width}. The dependence of $\Delta r$ on $R_{\rm bin}$ can be understood from how wide the energy spread is within the debris, as indicated by the scaling factor $(\Delta\epsilon / \epsilon)$. We attribute the larger $H$ at larger $R_{\rm bin}$ to weaker vertical gravity. If the debris are in vertical hydrostatic equilibrium, $dP/dz\simeq G\Mbh z \rho /R_{\rm bin}^{3}$. If density and pressure share the same scale height ($P dz/dP\simeq \rho dz/d\rho$) for a given $R_{\rm bin}$, and are virialized ($P/\rho\propto R_{\rm bin}^{-1}$), then $Pdz/dP\propto (P/\rho) R_{\rm bin}^{3}\propto R_{\rm bin}^{2}$. This scaling relation $P/\rho\propto R_{\rm bin}^{-1}$ is indeed found in our simulations (see the bottom panels of Fig.~\ref{fig:disk_evolution} and \ref{fig:disk_evolution_10x}).

For these spatial scales and their scalings, we can make two additional points: 1) The volume density scales as $R_{\rm bin}^{-5}$ because $\rho\simeq M_*/[2\pi R_{\rm bin}H\Delta r]$, consistent with the simulations. The peak density is $10^{-2}$ g cm$^{-3}$ for $R_{\rm bin}=10^{-4}\pc$, whereas it is $10^{-7}$ g cm$^{-3}$ for $R_{\rm bin}=10^{-5}\pc$. 2) The very small $H/\Delta r$ implies that photon diffusion is generally faster along the vertical direction.

While photons escape on the photon diffusion timescale $t_{\rm diff}$, a fraction of the disk would gradually fall toward the MBH and finally accrete over the viscous timescale $t_{\rm vis}$. If $t_\mathrm{visc}<t_\mathrm{diff}$, photons remain ``frozen'' into the debris and the luminosity is mostly determined by $t_\mathrm{visc}$. On the other hand, if $t_\mathrm{visc}>t_\mathrm{diff}$, photons diffuse out before the gas accretes, so the prompt luminosity at early times is determined by $t_\mathrm{diff}$. However, once the infalling debris begins accreting, the luminosity would arise from both accretion and radiative diffusion.

\begin{figure}
    \centering
    \includegraphics[width=0.45\textwidth]{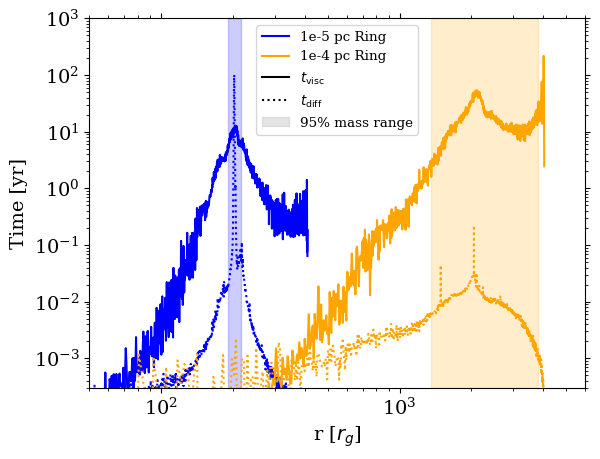}
    \caption{{ Diffusive cooling and viscous timescales (Eq.  \ref{eq:tdiff} and Eq. \ref{eq:tvisc} respectively) as a function of the radial distance} at $t\approx 5\mathcal{P}$ for both ring simulations. 
    Shaded regions enclose 95\% of the debris mass, highlighting the areas of greatest significance.
    $t_\mathrm{diff}$ peaks at the tori centers, corresponding to the densest regions.
    }
    \label{fig:ring_timescales}
\end{figure}

In Figure \ref{fig:ring_timescales}, we compare $t_{\rm diff}$ with $t_{\rm vis}$, averaged over radial shells and weighted by mass, throughout the simulated rings. We find that $t_\mathrm{visc}$ is two to three orders of magnitude longer than $t_\mathrm{diff}$ in both cases, which suggests that the prompt luminosity is primarily powered by the thermal energy radiated away, rather than by accretion.

\begin{figure}
    \centering
    \includegraphics[width=0.5\textwidth]{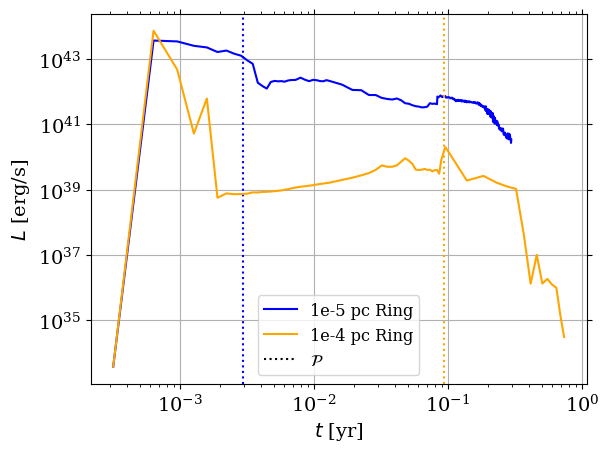}
    \caption{Bolometric
luminosity evolution of the TDEE in the two ring outcomes. The first simulation point has been manually set to $1L_\odot$ since the virtual grid does not properly resolve the star prior to disruption. 
The vertical dotted lines indicate the one period timescale for each of the two cases. The first peak corresponds to the micro-TDE, which is not well distinct from the encore in the $10^{-5}$~pc case.}
    \label{fig:ring_luminosities}
\end{figure}

\subsection{Luminosity}\label{subsec:ring_L}

Fig.~\ref{fig:ring_luminosities} shows the luminosity, estimated following the procedure explained in \S~\ref{Bolometric Luminosity from AREPO}, for the two simulations with $R_{\rm bin}=10^{-5}\pc$ (blue) and $10^{-4}\pc$ (orange). As in the direct encore scenario, the first peak corresponds to the micro-TDE by the sBH. 
We remind that, because our simulations neglect physical viscosity and the viscous timescale for accretion onto sBHs is relatively short, the diffusive sBH luminosity presented here may not accurately represent the actual sBH flare, which is likely dominated by accretion luminosity (see, e.g., \cite{Perets2016,Kremer2019,Wang2021}).

Following the initial sBH flare, the subsequent plateau in the luminosity arises from the TDEE ring. The luminosity for $R_{\rm bin}=10^{-5}\pc$ is approximately two orders of magnitude higher than that for $R_{\rm bin}=10^{-4}\pc$.
We can understand this dependence on
$R_{\rm bin}$ by deriving an approximate expression for the luminosity which relies on 
using the timescale considerations of the previous subsection. 
 The vertical photon diffusion can be approximated as $t_{\rm diff}\simeq \kappa \rho H^{2}/c$, where $\kappa$ is the Thomson opacity. The luminosity can then be estimated as
\begin{align}\label{eq:L_Rbin}
    L\simeq \frac{aT^{4}V}{t_{\rm diff}}\propto \frac{T^{4}}{\rho^{2} H^{2}}\propto R_{\rm bin}^{2},
\end{align}
where $V\simeq M_*/\rho$ is the volume of the debris. Here, we have used $T\propto R_{\rm bin}^{-1}$, $\rho\propto R_{\rm bin}^{-5}$, and $H\propto R_{\rm bin}^{2}$.  This scaling relation is consistent with our luminosity estimates from the simulations.

Based on our viscous timescale estimates (Fig.~\ref{fig:ring_timescales}), the luminosity may rise again to $\eta \ms c^{2}/t_{\rm vis}\simeq (1-5)\times 10^{43} (\eta/0.01)$ erg s$^{-1}$,  
due to the accretion onset after $\sim 10$ to 50 years since the micro-TDE, with the precise value depending on $R_{\rm bin}$.

\section{Discussion}\
\label{sec: discussion}

\subsection{Observational Implications}

Our hydrodynamical simulations of TDEEs with \arepo have allowed us to study in detail this special phenomenon, characterize its morphological features, and estimate the expected luminosity.

For direct encores, our simulations show that debris stream self-intersections can occur as early as one week after the micro-TDE, scaling as 
$(R_{\rm bin}/10^{-5}\pc)^{3/2}(\Mbh/10^{3}\Msol)^{-1/2}$.
This stage in the light curve corresponds to a sharp luminosity rise from shocks, observable in the optical/UV band (see Fig.~\ref{fig:direct_luminosities} \textbf{in Appendix \ref{sec:SEDs}}). While the folding of debris streams into self-intersections near the MBH is complex and highly sensitive to initial conditions, the peak luminosity remains robust against these variations.

X-ray emission is expected both from the initial micro-TDE (e.g., \citep{Perets2016}) and from subsequent accretion onto the MBH on the viscous timescale, with the X-ray fraction of the accretion luminosity generally increasing for lower-mass MBHs.
The micro-TDE luminosity is roughly four orders of magnitude above the Eddington limit of a $10\Msol$ sBH, which is likely to drive strong outflows and winds, making this initial event possibly appear like an X-ray rich, long-duration gamma-ray burst \citep{Perets2016}. 
We find that for both morphologies of TDEEs, the two X-ray signatures 
are separated in time approximately by months to years. 

Here we bring up the recent observations of AT 2023lli by \citet{Huang2024} which show an optical/UV bump 2 months prior to the TDE's peak. Their observations show the early bump is two orders of magnitude smaller than the eventual global peak. This appears to align well with the peak interval and luminosity corresponding to a $\approx10^6\Msol$ TDEE where the micro-TDE is Eddington limited. In fact, their work also infers that the central BH mass is $10^{6.4\pm0.47}\Msol$. In the perfectly circular sBH-MBH case, their separation would be $R_\mathrm{bin}\approx4 \times 10^{-4}\pc$ based on the time difference between the bump and the global peak of $50\text{ days}$. Our results suggest that a TDEE of this configuration would have a encore luminosity peak at $\approx10^{41}\text{ erg/s}$.

Additionally, the delayed X-ray peak of AT 2023lli may be explained by the dynamics of TDEEs. 
Because the debris
initially lie farther from the MBH than that in ordinary TDEs, the formation of an accretion disk—and thus the onset of X-ray emission—may be delayed due to an extended viscous timescale. The $\sim$150-day lag between peak optical and peak X-ray
luminosity in AT2023lli is consistent with the range expected from the
viscous timescale of the debris structure’s inner edge. While it is
not yet certain that AT~2023lli is a TDEE, its properties align with
several predicted features and merit further study.

\subsection{Encores as IMBH Probes}
\label{sec: discussion IMBH}
The existence of IMBHs in NSCs remains tentative, yet nuclear TDEs provide promising probes for their detection. Whereas most of the prior work of \citet{Ryu2024TDEE} has concentrated on estimating
TDE rates from IMBHs, the direct encore flares predicted here -- lower luminosity in X-rays, followed by higher luminosity peak in optical/UV -- offer an independent diagnostic  for constraining IMBH masses in NSCs. 

Since encore debris circularization begins at roughly the freefall timescale $t_\mathrm{ff}$, the time delay between the micro-TDE and TDEE flares encodes information about $\Mbh$ and $R_\mathrm{bin}$, offering  potential insights into the presence of IMBHs in NSCs.  
Furthermore, because both TDEE rates and their electromagnetic signatures depend on 
$R_\mathrm{bin}$ \citep{Ryu2024TDEE},  their detection and characterization can be also informative of the properties of the innermost sBH population of the NSC, which is generally hard to probe by other means.

The time delay between the initial micro-TDE flare and its encore is very short, on the order of $\sim$ week timescale for the specific IMBH mass ($10^3 M_\odot$) and binary separation ($10^{-5}$~pc) simulated here.  It becomes longer for larger binary separations ($\propto R_{\rm bin}^{3/2}$), while it is shortened by larger central masses 
($\propto M_{\rm MBH}^{-1/2} $).
Because micro-TDEs are expected to be most prominent in high-energy X-rays/$\gamma$-rays, we encourage systematic follow-ups and cross-correlations of nuclear transients across multiple wavebands.

\subsection{General Relativistic Effects}

Because most debris in our simulations for  the encore flare remains sufficiently far from the MBH, the inclusion of relativistic effects is not expected to impact the overall evolution. The primary relativistic effect on tidal streams would be periapsis advance of the debris, given by $\Delta \phi \simeq$ 1 radian $(r_{\rm p}/10r_{\rm g,MBH})^{-1}$ for highly eccentric geodesics \citep{Hobson+2006}, where $r_{\rm p}$ is the pericenter distance.
For most simulations, we find that $\Delta\phi < 10^{-3}$ radians relative to both BHs; however, certain debris trajectories ($\pi/2\leq \theta \leq3\pi/4$) for our direct encore simulations have $10^{-2}<\Delta\phi < 10^{-1}$. In both cases, the expected relativistic effect is still small and thus expected to have a minor contribution in decreasing the circularization time of debris via stream self-crossing. From this post-process test, we therefore conclude that a Newtonian model for gravity is sufficient to characterize the observables predicted in this paper. However, relativistic effects eventually become important when the debris lose its orbital energy via shocks and accumulate in the vicinity of the MBH. 

While our simulations are constrained to Newtonian mechanics, we can analytically predict the gravitational waves (GWs) emitted by the circular orbit of the  sBH-MBH binary. 
With reduced mass $\mu\approx \Mbh$, chirp mass $\mathcal{M}\approx \Mbh^{2/5}\sbh^{3/5}$, and distance to the observer $R_\mathrm{obs}$, we derive that the GW strain  from TDEE sources is

\begin{equation}
\begin{split}
    h\sim 2.3 \times10^{-23}\left(\frac{\sbh}{10\Msol}\right) \left(\frac{\Mbh}{10^6 \Msol}\right) 
    \\ \times
    \left(\frac{R_\mathrm{bin}}{10^{-5}\pc}\right)^{-1} \left(\frac{R_\mathrm{obs}}{400\Mpc}\right)^{-1}
\end{split}
\end{equation}
and the frequency of the waves $f_\mathrm{GW}=2/\mathcal{P}$ is
\begin{equation} 
f_\mathrm{GW}= 2.1 \times 10^{-5} \mathrm{\;Hz} 
\left(\frac{\Mbh}{10^{6}\Msol}\right)^{1/2}
\left(\frac{R_\mathrm{bin}}{10^{-5}\pc}\right)^{-3/2}\,.
\end{equation}

In general, the expected strain from extreme mass ratio inspirals (EMRIs) forming TDEEs is too small to be detected corresponding to direct encores, and ring encores are tidally limited by the MBH. For the planned 
 Laser Interferometer Space Antenna
(LISA) sensitivity, most EMRIs would be outside expected sensitivities \citep{Robson2019}. However, event rates from TDEEs offer valuable insight into BH populations in NSCs.

Inverting the equation for rates from \citet{Ryu2024TDEE}, we can obtain a lower bound on the radial distribution of sBHs in NSCs. A measurement of $d\dot N_{\rm TDEE}/dR_\mathrm{bin}$ in the near vicinity of a MBH is possible via the use of TDEE's significant and measurable dependence on orbital period. Coupled with estimations of the mass of the MBH at the centre of a given galaxy such as those done by \citet{Huang2024}, values of $R_\mathrm{bin}$ can be extracted from light curve data.

Given a population of detections, we can then calculate
\begin{equation}
    \frac{dN_{\rm sBH}}{dR_\mathrm{bin}} = \frac{d\dot N_{\rm TDE}}{dR_\mathrm{bin}}(n_\star(R_\mathrm{bin}) v(R_\mathrm{bin}) \Sigma(R_\mathrm{bin}))^{-1}\,,
\end{equation}
where $n_\star$ is the stellar number density, $v$ is the maximal velocity, and $\Sigma$ is the encounter cross section.
While the range of distances for which TDEEs are more likely to occur is immediately unresolvable for upcoming space based missions, the BHs will eventually evolve into the LISA band as they radiate gravitational energy.

Meanwhile, the entire population of these unresolvable binaries will sum together and contribute to the stochastic GW background of space-based detectors. The background bolometric GW luminosity predicted by TDEE sources can be estimated as
\begin{equation}
\begin{split}
    L_\mathrm{GW, \,background} = \int dR_\mathrm{bin}dM_\mathrm{sBH}\\
    \left[\frac{dN_\bullet}{dR_\mathrm{bin}} \phi(M_\mathrm{MBH}) L_\mathrm{GW}(M_\mathrm{MBH},R_\mathrm{bin})\right]\, 
\end{split}
\end{equation}
where $\phi(M_\mathrm{MBH})$ is the initial mass function of the central MBH. Although a single EMRI in the TDEE $R_\mathrm{bin}$ range is not detectable, their sum may be appreciable.

\section{ Conclusions and Outlook}
\label{sec: Conclusion}

In this work, we have carried out the first hydrodynamical simulations
of tidal disruption encore events using the moving-mesh code {\tt AREPO}. Our study focused on the disruption of Sun-like stars by
stellar-mass BHs  within the potential of a more massive
BH, exploring the subsequent evolution of debris into
encore flares. We identified two primary classes of encores: {\em direct
encores}, in which debris streams plunge toward the MBH and produce
shock-powered flares, and {\em ring encores}, in which debris circularize
at larger radii, leading to delayed accretion and extended emission.
The time delay between the primary flare from the micro-TDE onto the
sBH and the encore onto the MBH is determined by the freefall timescale
in the former case and the orbital period in the latter.

For the range of parameters studied here,
we find that the delay between the micro-TDE and the encore
flare can range from weeks to several decades, depending sensitively
on the MBH mass and the sBH–MBH separation. The resulting light curves
exhibit characteristic double-flare structures: a primary flare due
to the micro-TDE, likely brighter in X-rays/$\gamma$-rays,
followed by  optical/UV emission 
powered by debris self-intersections, with contribution 
from viscous accretion onto the MBH. The latter, especially if the MBH
is an IMBH, will likely emerge in X-rays.
In some regimes, the
accretion luminosity can exceed the Eddington limit for $M_{\rm
MBH}\sim10^3 $M$_\odot$, raising the possibility of strong outflows or
jet activity. These features position TDEEs as a novel class of
nuclear transients, potentially explaining observed candidate
TDE events with non-standard temporal decays and delayed rebrightenings.

Looking forward, several directions merit further
exploration. Extending the simulations to a broader range of stellar
masses and internal structures will elucidate how TDEE energetics and
timescales scale with stellar properties. Incorporating radiation
transport will provide more reliable predictions for spectra and
observational diagnostics. Embedding our framework within cluster
dynamical models will allow population-level predictions for TDEE
rates and their connection to nuclear star cluster structure. From an
observational perspective, wide-field optical/UV surveys (e.g., Rubin, ULTRASAT) combined with
sensitive X-ray follow-up (e.g., eROSITA, Einstein Probe, Athena) are well
positioned to identify candidate TDEEs through their distinctive
double-flare light curves. Finally, in cases where sBHs inspiral
toward MBHs, TDEEs may occur contemporaneously with extreme mass ratio
inspirals detectable by LISA, offering a multi-messenger pathway to
probe the dynamics of dense nuclear star clusters.

In summary, TDEEs open a new observational window onto the
astrophysics of dense stellar environments. In particular,
encore flares may provide an independent diagnostic for the long-sought
population of intermediate-mass BHs in nuclear star
clusters. As both time-domain surveys and gravitational-wave
observatories advance, TDEEs may emerge as a complementary probe of BH demographics and strong-gravity astrophysics.

\section*{Acknowledgements}
\label{sec: Acknowledgements}

R.P. and I.J. gratefully acknowledge support by NSF award AST-2006839 and NASA award 80NSSC25K7554. The authors would also like to thank Stony Brook Research Computing and Cyberinfrastructure, and the Institute for Advanced Computational Science at Stony Brook University for access to the SeaWulf computing system, made possible by grants from the National Science Foundation (\#1531492 and Major Research Instrumentation award \#2215987), with matching funds from Empire State Development’s Division of Science, Technology and Innovation (NYSTAR) program (contract C210148).

\section{Appendix A: Resolution Tests}
\label{sec: res tests}

Our simulations are run with a stellar model of 500k cells initially. This resolution was chosen after a series of simulations of increasing resolutions were preformed until we had demonstrated the convergence of post-process values including luminosity and the relevant timescales.

\begin{figure}
    \centering
    \includegraphics[width=.9\linewidth]{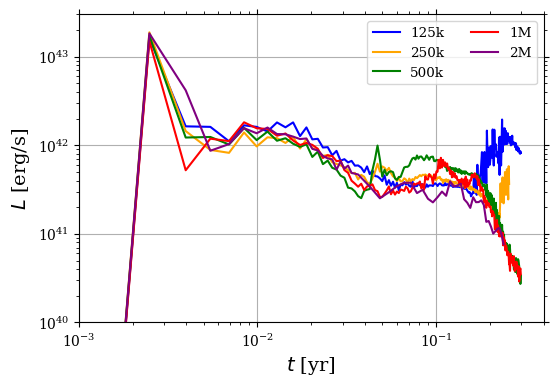}
    \caption{Resolution test: Luminosity for the ring simulation with $R_{\rm bin}=10^{-5}$~pc with relaxed \texttt{MESA} model stars of varied initial number of cells. We adopted the 500k resolution since values converge on the simulated timescale. }
    \label{fig:TDEE_Ring_Resolution_Luminosities}
\end{figure}

As an illustrative example, Fig.~\ref{fig:TDEE_Ring_Resolution_Luminosities} shows the results of our resolution test for one of the models, the ring-outcome simulation with
$R_{\rm bin}=10^{-5}$~pc.
The figure presents the predicted luminosities obtained from relaxed \texttt{MESA} stellar models with initial mesh resolutions ranging from 125k to 2M cells. The results demonstrate that a resolution of 500k cells is sufficient, as it yields converged luminosities consistent with those at higher resolutions while minimizing computational cost.

\section{Appendix B: \mesa Stellar Model}

\begin{figure*}
    \centering
    \includegraphics[width=\textwidth]{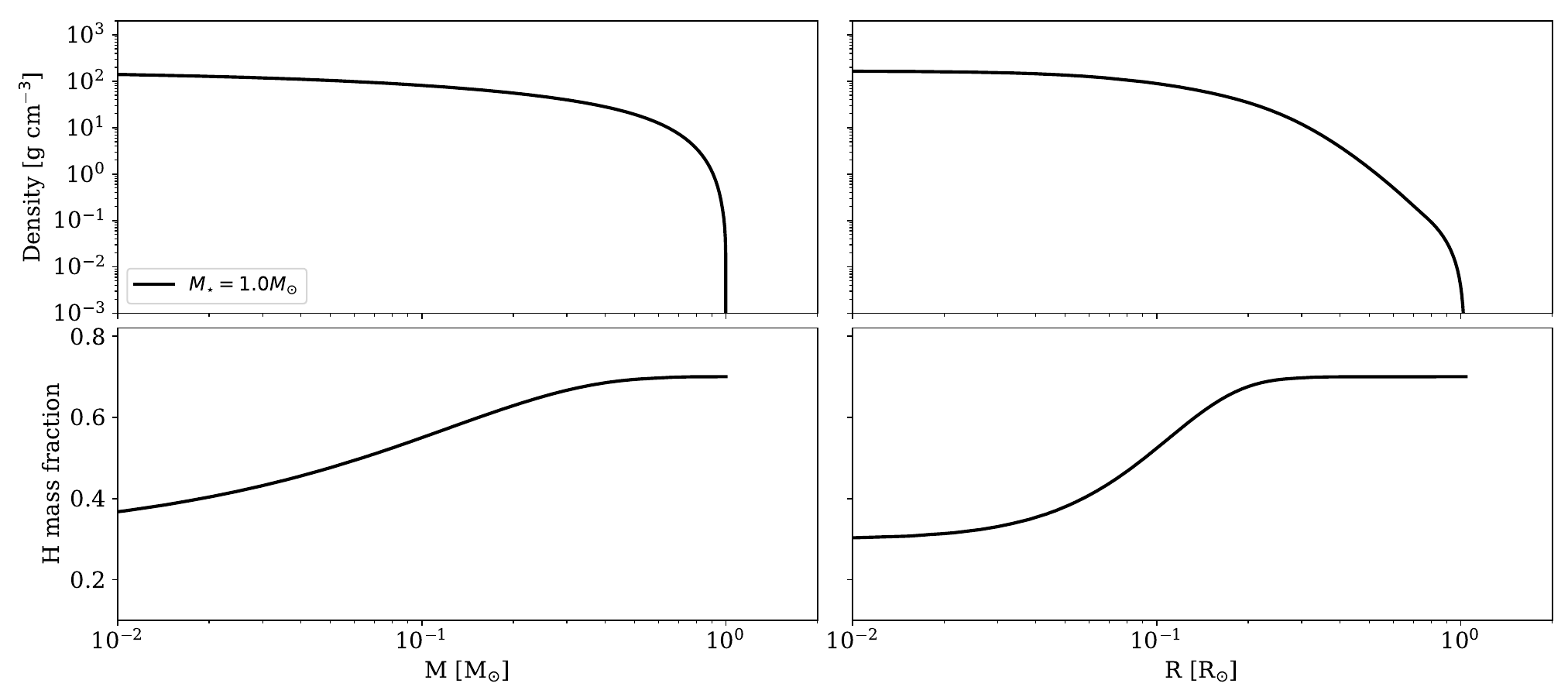}
    \caption{\label{fig:initialprofile} Density and hydrogen mass fraction profiles of the $1\Msol$ model consider. The left (right) panels depict the profile as a function of enclosed mass (radius).    }
\end{figure*}

As a reference, Fig.~\ref{fig:initialprofile} shows the density and hydrogen mass fraction profiles of the
$1M_\odot$ star modeled with \texttt{MESA} and used in our simulations.

\section{{ Appendix C: Spectral Properties of TDEEs}}
\label{sec:SEDs}
{
Using the upward flux $F_\mathrm{rad}(r,\phi)$ described in section \ref{Bolometric Luminosity from AREPO}, together with the temperature at the photosphere along the virtual grid $T_\mathrm{ph}(r,\phi)$, we can construct  approximated spectral energy distributions (SEDs) associated with different TDEE morphologies. 

We discretize the photosphere into surface elements indexed by $k$, each characterized by a photospheric temperature $T_{\mathrm{ph},k}$ and an associated bolometric luminosity contribution $L_k$, obtained by integrating the upward radiative flux over the corresponding area element,
\begin{equation}
L_k \equiv \int_{\Delta A_k} F_{\mathrm{rad}}(r,\phi)\, dA .
\end{equation}

To construct a continuous approximation to the emergent spectrum, we bin the photospheric temperatures into a set of 500 temperature bins $\{T_i\}$ with centers $T_i$. The luminosity associated with each temperature bin is computed as 
\begin{equation}
L(T_i) = \sum_{k \in i} L_k\,,
\end{equation}
where the sum runs over all surface elements $k$ whose photospheric temperatures fall within the $i$-th temperature bin.

Each temperature bin is assumed to emit as a blackbody at temperature $T_i$. The specific luminosity contributed by a given bin is therefore
\begin{equation}
L_{\nu,i} = L(T_i)\,
\frac{B_\nu(T_i)}{\sigma T_i^4},
\end{equation}
where $B_\nu(T)$ is the Planck function and $\sigma$ is the Stefan–Boltzmann constant. 

Although we normalize by $\sigma T_i^4$ so that integrating $L_{\nu,i}$ over frequency should recover the bolometric luminosity $L(T_i)$, we find that the integral of our SEDs does not exactly retrieve the bolometric luminosities in Section \ref{Bolometric Luminosity from AREPO}. This discrepancy is due to variations in the calculation of $L(t)$ from the radiative energy contained in a column of the virtual grid (\ref{eq: Frad}) versus a calculation of $L(t)$ from solely the photosphere temperature. For the results provided in the main text, we avoided calculating the bolometric luminosity from $T_\mathrm{ph}$ to mitigate \arepo's incomplete radiation transfer capabilities, but this agnostic approach is not possible for spectral analysis. Thus, we present these findings with caution since our simulations do not include radiation transfer and the virtual grid does not resolve the photosphere with high precision; however, we believe these estimates can still be useful to broadly categorize the TDEE spectral observability.

The total spectral energy distribution is then obtained by summing over all temperature bins,
\begin{equation}
L_\nu = \sum_i L(T_i)\,
\frac{B_\nu(T_i)}{\sigma T_i^4}.
\end{equation}
This procedure is repeated for several simulation times, to construct a sequence of SEDs that illustrate the spectral evolution of different TDEE morphologies (Figures \ref{fig:SED_135}-\ref{fig:SED_ring_10x}).

In Figure \ref{fig:SED_135}, we show the SED results for our fiducial choice of direct encore for the case $\theta=3\pi/4$. Before the start of the encore, our simulations indicate a shift to lower frequencies and luminosities from the initial micro TDE. As the encore begins around $t=1.7\;\rm{days}$, the spectral peak broadens toward higher frequencies and subsequently increases in luminosity. As time progresses beyond the onset of the encore, the peak luminosity decreases again. This spectral shift is explained by encore shocks causing a secondary peak at $\nu\sim 5 \times 10^{15} \;\mathrm{Hz}$ whose luminosity increases between $t=1.7$ and $t=7.3$ days as the shocks persist. As this encore peak grows, the initial peak from the micro TDE continues to cool into the optical band, so that by $t=7.3$ the encore peak becomes dominant.  For $t>7.3$ days 
the debris undergoes adiabatic cooling,
although specific dynamic idiosyncrasies may imprint minor features on the subsequent evolution of the SED.

\begin{figure}
    \centering
    \includegraphics[width=.45\textwidth]{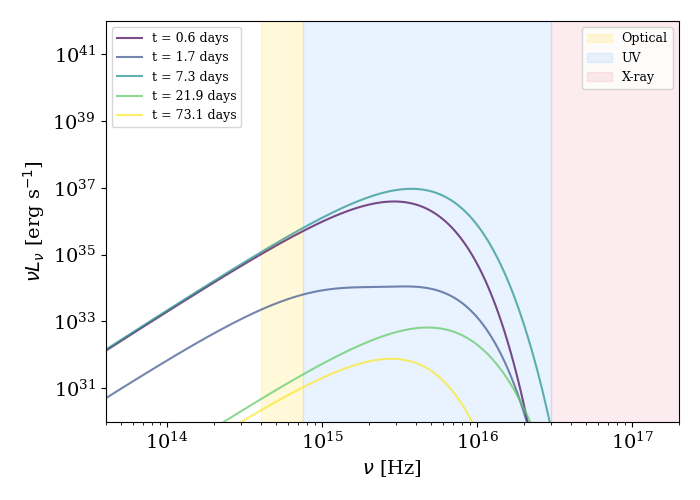}
    \caption{{ SED  for the direct encore case of Figure \ref{fig:direct encore evo}, showed at five different times in order to illustrate the temporal evolution of the SED. 
    Colored regions indicate approximate boundaries for optical (yellow), UV (blue), and X-ray (red) bands.  }}
    \label{fig:SED_135}
\end{figure}

For other direct encore configurations, we note that the spectral evolution follows a similar trend of peak widening toward higher frequencies and then increasing peak luminosity; however, specific hydrodynamic effects can imprint observable differences on the SED evolution. 

In the ring encore scenario, the temporal
 evolution of the SED is comparatively simpler. In Figure \ref{fig:SED_ring}, we show SEDs for the $R_\mathrm{bin}= 10^{-5}\pc$ ring encore. For this morphology, encore shocks shift the peak toward higher frequencies. As the shocks associated with disk/torus formation continue after approximately one orbital period, adiabatic expansion cools the debris culminating in the SED peak shifting toward lower frequencies.

\begin{figure}
    \centering
   \includegraphics[width=.45\textwidth]{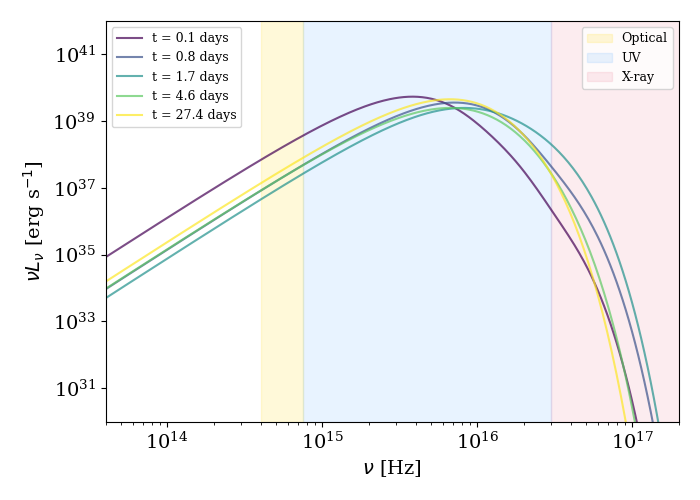}
    \caption{{ SED corresponding to the temporal evolution of the encore of Figure \ref{fig:disk_evolution}. Plotting conventions are the same as in Figure \ref{fig:SED_135}; however, because this case corresponds to
     a ring encore with orbital period $\mathcal{P}\approx 1.1 \mathrm{\;days}$, the spectral evolution is markedly different.}}
    \label{fig:SED_ring}
\end{figure}

When encores happen in less gravitationally bound cases like the $R_\mathrm{bin}= 10^{-4}\pc$ ring encore shown in Figure \ref{fig:SED_ring_10x}, cooling can already be significant before encore onset, similarly to the direct encore case. This makes the encore more pronounced, as the spectral peak luminosity 
exhibits larger fluctuations on
month-long timescales.

\begin{figure}
    \centering
    \includegraphics[width=.45\textwidth]{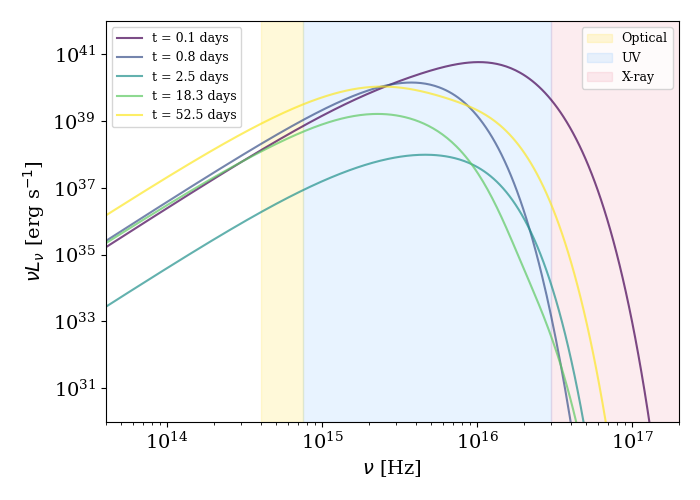}
    \caption{{ SED corresponding to the temporal evolution of the encore of Figure \ref{fig:disk_evolution_10x}. The SED trends are similar to those of Figure \ref{fig:SED_ring_10x}, but the longer orbital period ($\mathcal{P}\approx 34 \mathrm{\;days}$) allows for more cooling before the onset of shocks in the encore.}}
    \label{fig:SED_ring_10x}
\end{figure}

While these SEDs are order-of-magnitude estimates, our results highlight TDEEs as compelling targets for multi-wavelength astronomy, 
as distinct structures can produce spectral peaks in the optical, UV, and X-rays. Moreover, temporal variability in the spectral emission may provide a novel avenue for probing the astrodynamics and population properties of nuclear star clusters. 
}

\bibliography{biblio}

\end{document}